\documentstyle[aps,epsf,epsfig,preprint]{revtex}
\def\prl#1#2#3{{ Phys. Rev. Lett.} {\bf #1}, #2 (#3)}

\def\pla#1#2#3{Phys. Lett. A {\bf #1}, #2 (#3)}

\def\pra#1#2#3{Phys. Rev. A {\bf #1}, #2 (#3)}
\def\prb#1#2#3{Phys. Rev. B {\bf #1}, #2 (#3)}

\def\jpa#1#2#3{J. Phys. A {\bf #1}, #2 (#3)}
\def\ijbc#1#2#3{Int. J. Bifurcation and Chaos {\bf #1}, #2 (#3)}
\def\cmp#1#2#3{Comm. Math. Phys.  {\bf #1}, #2 (#3)}

\def\physd#1#2#3{Physica D {\bf #1}, #2 (#3)}

\def\gm{\mbox{g}}
\def\ecr{$\epsilon_c$}
\def\eps{$\epsilon$}

\def\eg{e.~g.}

\def\beqr{\begin{eqnarray}}
\def\eqnr{\end{eqnarray}}
\def\beq{\begin{equation}}
\def\bc{\begin{center}}
\def\ec{\end{center}}
\def\eqn{\end{equation}}
\begin{document}
\title{Critical States and Fractal
Attractors in Fractal Tongues:\\ Localization in the Harper map}
\author{Surendra Singh Negi and Ramakrishna Ramaswamy} 
\address{School of Physical Sciences\\ Jawaharlal Nehru
University, New Delhi 110 067, INDIA}
\date{\today}
\maketitle
\begin{abstract}
Localized states of Harper's equation correspond to strange
nonchaotic attractors (SNAs) in the related Harper mapping. In
parameter space, these fractal attractors with nonpositive Lyapunov
exponents occur in fractally organized tongue--like regions which
emanate from the Cantor set of eigenvalues on the critical line
$\epsilon = 1$. A topological invariant characterizes wavefunctions
corresponding to energies in the gaps in the spectrum. This permits a
unique integer labeling of the gaps and also determines their scaling
properties as a function of potential strength.
\end{abstract}
The Harper equation \cite{harper},
\beq
\psi_{n+1} + \psi_{n-1} + V(n) \psi_n = E \psi_n  \label{harpq},
\eqn
where $\psi_n$ denotes the wave-function at lattice site $n$, and
$V(n) = 2\epsilon \cos 2 \pi (n \omega + \phi_0)$ with $\omega$
irrational has been extensively studied in the context of
localization.  This discrete Schr\"odinger equation for a particle in
a quasiperiodic potential on a lattice, arises in a number of
different problems \cite{hofstadter,contexts,geisel}.  It is known
\cite{aubry} that the eigenstates can be extended ($\epsilon < 1$),
localized ($\epsilon > 1$), or critical ($\epsilon = 1 \equiv$ \ecr),
when the eigenvalue spectrum is singular--continuous
\cite{bellisard}, and the states are power--law localized
\cite{kohmoto,pandit}. Renormalization group studies
\cite{op,kspla,sk} of this model have been very effective in
establishing the multifractal nature of the wavefunctions of such
states and of the eigenvalue spectrum \cite{kohmotoa}.

On transforming to Ricatti variables \cite{luck} $\psi_{n-1}/\psi_n 
\to x_n$, Eq.~(1) reduces to the (equivalent) Harper map \cite{ks}, 
\begin{eqnarray}
\label{harpc}
x_{n+1} &=& -[x_n - E + 2\epsilon \cos 2\pi \phi_n]^{-1} \\
\phi_{n+1} &=& \{\omega+\phi_n\} 
\label{harpcp}
\end{eqnarray}
where $\{y\} \equiv y$ mod 1.  Viewed as a skew--product dynamical
system, this is now a driven mapping of the infinite strip
$(\-\infty,\infty)\otimes [0,1]$ to itself. Irrational $\omega$
implies that the forcing in Eq.~(\ref{harpc}) is quasiperiodic, the
lattice site index $n$ in the quantum problem becoming the iteration
or time index in the map. 

The Harper map, which we study in this Letter, provides an alternate
means of analyzing the eigenvalue spectrum of the Harper equation.
Boundary conditions that must be imposed on Eq.~(\ref{harpq}) in
order to determine eigenstates become conditions on the dynamical
states in the map, Eqs.~(2-3), where $E$ now appears as a parameter.
Note that since the map is reversible, there is no chaotic motion,
and furthermore, because the driving is quasiperiodic, there are also
no periodic orbits.  For large enough \eps, the attractor of the
dynamics is a fractal, and on this attractor, the dynamics is
nonchaotic \cite{ks,prss}: these are therefore strange nonchaotic
attractors (SNAs) \cite{gopy} which are generic in
quasiperiodically forced systems. In addition there can be a variety
of other nonfractal quasiperiodic (torus) attractors.

When $E$ is an eigenvalue, a correspondence relates localized states
of the quantum problem to SNAs in the associated map. This
equivalence that was first noted by by Bondeson, Ott and Antonsen
\cite{bondeson} in a study of the continuous version of the
same problem.  The wavefunctions for critical and extended states
have a quasiperiodic symmetry \cite{prss}, while the fractal
fluctuations of the amplitudes of the localized states
\cite{sk} appear as fractal density fluctuations in the attractors of
the map.  The nontrivial Lyapunov exponent of the system
is given by
\beq
\lambda = \lim_{N \to \infty} {1 \over N} \sum_{i=1}^N y_n,
\label{le}
\eqn
where $y_n = \ln x_{n+1}^2$, the so--called stretch exponent, is the
derivative of the mapping, Eq.~(\ref{harpc}).
At \ecr, if $E$ is an eigenvalue of the
quantum problem, then this quantity is exactly zero
\cite{prss}; see Fig.~1.  Above \ecr, the localization length for
quantum states, $\gamma$, is inversely related to the Lyapunov
exponent \cite{aubry}, $\lambda^{-1} = -\gamma/2$. Knowledge of this
equivalence thus permits the complete determination of the quantum
spectrum of the Harper system through a study of the Lyapunov
exponents of the Harper map.  At every eigenvalue (Fig.~1) there is a
bifurcation from a quasiperiodic attractor to a SNA, when the
Lyapunov exponent becomes zero \cite{hof}.

The spectrum of the Harper equation is invariant under the
transformation $\omega \to 1-\omega$, and is symmetric about $E=0$,
so it suffices to consider only positive eigenvalues and $\omega >
1/2$. The behaviour of the spectral {\it gaps} has been of
considerable interest \cite{geisel}, and we study this here by
describing the phase--diagram of this system for $\omega_{\gm} =
(\sqrt 5 -1 )/2$, the inverse golden--mean ratio.

At $\epsilon = 0$, the states of the quantum system form a band
between energies $0 \le E \le 2$. As \eps~ is increased, the gaps
open up and merge as $\epsilon \to \epsilon_c$, giving a singular
continuous spectrum. Below \ecr, when the states are extended the
dynamics of the classical system is on ``three--frequency''
quasiperiodic orbits \cite{bondeson} with Lyapunov exponent equal to
0. Above \ecr, localized states correspond to SNAs with negative
Lyapunov exponent (see Fig.~2a for an example), while critically
localized states at \ecr~are exceptional and correspond to SNAs with
a zero Lyapunov exponent \cite{prss}. For energies in the gaps,
$\lambda < 0$, and the motion is on two--frequency quasiperiodic
(1--dimensional) attractors (an example is shown in Fig.~2b), which
wind across the $(x,\theta)$ plane an integral number of times.
Wavefunctions of Eq.~(\ref{harpq}) at these energies do not satisfy
the appropriate boundary conditions and are non--normalizable.  The
number of windings of the corresponding attractor, $N$ is a
topological invariant for {\it all} orbits in the gaps. This integer
index for each gap \cite{winding} counts the number of changes of
sign (per unit length) of the wavefunction, and is thus related to
the integrated density of states (IDS) \cite{numbertheory}. The
gap--labelling theorem \cite{numbertheory} states that each gap can
be labeled by the value that the IDS takes on the gap; in the Harper
system, this is also the winding number \cite{winding}, and (for $E
\ge 0$) on the gap labeled by the index $N$, this takes the value
\beq
\Omega_N (E) = \mbox{max}(\{ N \omega \},1-\{ N \omega \}).
\label{winding}
\eqn
(The symmetrically located gap with $E \le 0$ with index $N$ has
winding number $\Omega_N (E) = \mbox{min}(\{ N \omega \},1-\{ N
\omega \})$).

There is thus a 1--1 correspondence between the gaps and the
integers.  Furthermore, since the IDS is a continuous nondecreasing
curve, it is possible to specify the gap ordering: this depends on
the continued fraction representation of $\omega$. This latter
problem has been studied earlier by Slater \cite{slater}, and is also
encountered in the context of level statistics of two--dimensional
harmonic oscillator systems \cite{pbg,pr}.  Consider the set
of numbers $y_j = \{j\omega\}, j= 1, 2,\ldots, m$. For {\em any}
$\omega$ and any $m$, it has been shown \cite{slater,pbg} that an ``ordering
function'' can be defined, giving a permutation of the indices, $j_1,
j_2,\ldots,j_m$, such that $y_{j_i} \le y_{j_k}$ if $i<k$. This
result can be directly adapted to the present problem so as to obtain 
the complete ordering of gap labels with $E$ \cite{reorder}.

The resulting structure of the gaps can be described via a simpler 
construction for the case of $\omega = \omega_{\gm}$. Recall that
$\omega_{\gm}= \lim_{k\to\infty} F_{k-1}/F_k$, where the Fibonacci 
numbers $F_k$ are defined by the recursion $F_{k+1} = F_{k-1} + F_k,$ 
with $F_0 = 1, F_1 = 2$.  Now consider a Cayley tree, arranged as shown in
Fig.~3, with each node (except the origin, labeled 0) having two
successors. Nodes at the same horizontal level are at the same
generation.  The rightmost node at each generation is labeled by
successive Fibonacci integers, while the leftmost are half the
successive even Fibonacci integers. The non--Fibonacci numbers are
then identified with nodes as follows: for given label $m$, the
parent node $i_m$ is the smallest available such that the sum
$(i_m+m)$ is a Fibonacci number. The sub--tree rooted at node $F_k$
contains a sequence $F_jF_{j+k}, j=1,\ldots$ which are placed
alternately to the left and right; this suffices in determining the
placement of all other integers within that subtree \cite{nr2}.

Every pair of integers, $i_1$ and $i_2$, with $i_2 > i_1$, has
two possibilities as to how they are relatively placed on this
graph. Either 
\begin{enumerate}
\item
$i_1$ is an ancestor of $i_2$, i.e. there is a directed
path connecting $i_2$ to $i_1$. If this path is to the left
at node $i_1$, then $i_2 \prec i_1$. (If to the right, 
then $i_1 \prec i_2$.)\\
\noindent or

\item
$i_0$ is the most recent common ancestor of $i_1$ and $i_2$. If the 
path from $i_0$ to $i_1$ is on the left at $i_0$, then
$i_1 \prec i_2$. (Similarly, if it is to the right, then
$i_2 \prec i_1$.)
\end{enumerate}
This gives a unique ordering of the integers (see Fig. 3) with
the relation $\prec$ being transitive (if $i \prec j$ and $j \prec m$
then $i\prec m$),
$$\ldots \prec 4 \prec  \ldots \prec 9 \prec \ldots \prec 1 
\prec \ldots \prec 7 \prec \ldots \prec 2 \prec \ldots \prec 
11 \prec \ldots \prec 21 \prec \ldots \prec 0.$$

The gaps appear in {\it precisely} this order: if $k \prec 
\ell$, then gap $k$ precedes gap $\ell$ in the positive energy 
spectrum of the critical Harper map (Fig.~1b). 
Following the procedure which is described in detail in
\cite{pbg,pr}, similar Cayley trees can be constructed for any 
other irrational frequency.  For each $\omega$, depending on its 
continued fraction representation, there is a unique 
reordering of the integers corresponding to the ordering 
of the gaps.

Each gap is further characterized by its width, $w_m$ and by its
depth $d_m$ both of which are functions of \eps.  The depth has no
obvious quantum--mechanical interpretation, $-d_m$ merely being the
minimum value that the Lyapunov exponent takes in the $m$th gap, and
it decreases with order $d_m > d_n$ if $m< n$, scaling, at \ecr~ as
$d_N \sim 1/N$ (see Fig.~4a). The behaviour of the gap widths is more
complicated and depends on the details of the Cayley tree.  These are
nonnmonotonic as a function of gap index, but come in families: gaps
belonging to a given family scale as a power, $w_N
\sim 1/N^{\theta}$. The fastest decreasing are the Fibonacci
gaps, 1,2,3,5,8,\ldots,$F_k$,\ldots ($\theta \equiv \theta_r
\approx 2.3$), while the slowest is the family
1,4,17,\ldots,$F_{1+3k}/2$,\ldots ($\theta \equiv \theta_l \approx
1.88$): these are respectively the successive rightmost and leftmost
nodes on the Cayley tree in Fig.~3 (see Fig.~4b). Other families,
which can be similarly defined on subtrees, also obey scaling, with
exponents between $\theta_l$ and $\theta_r$. When the gaps are
ordered by rank $r$, then they scale as $w_r \sim 1/r^2$: this is
consistent with the previously (numerically) obtained \cite{geisel}
gap distribution $\rho(s) \sim s^{-3/2}$, which has also been derived
exactly through the Bethe ansatz \cite{wiegman}.

Above \ecr, the states are exponentially localized. For all localized
states, irrespective of energy, the localization length or Lyapunov
exponent is identical \cite{aubry}. The gaps which dominate the
spectrum at \ecr, persist for larger \eps, but decrease in width
according to the (empirical) scalings (see Fig.~4b)
\beqr
\label{sc1}
w_N &\sim& {1 \over {N^{\theta} \epsilon^{N-1}}}\\ d_N &\sim& {1
\over {N \epsilon^N}},
\label{scal}
\eqnr
(where $\theta$ is particular to the family to which the gap
belongs).  

The dynamics of the Harper map corresponding to localized
states is on SNAs \cite{ks}, while that in the gaps continues to be
on 1--dimensional attractors similar to those below \ecr. However,
since the gaps decrease in width, most of the dynamics is now on
SNAs. By continuity, therefore, the SNA regions must start at each
eigenvalue at \ecr, and widen gradually since for large \eps~ the
spectrum lies in the range $0 \le E \le 2\epsilon$.  A phase--diagram
for this system in the $E-\epsilon$ plane is shown schematically in
Fig.~5.  The dynamics is entirely on fractal attractors with a
negative Lyapunov exponent in the tongue--like regions, each of which
starts at an eigenvalue at \ecr. The fractal (Cantor set) spectral
structure is thus reflected in the hierarchically organized fractal
``tongues''.

The equivalence between the Harper equation and the Harper map thus
provides a new mode of analysis of this problem which arises in
numerous contexts
\cite{harper,hofstadter,contexts,geisel,aubry,bellisard,kohmoto,pandit,op}.
The singular continuous nature of the eigenvalue spectrum, which has
been the subject of considerable theoretical study,
has been detected
in experiments \cite{experiments} as well, and therefore an
understanding of the gap widths and their variation with energy and
potential strength is of importance.

The present technique gives a simple but powerful method for the
study of the spectrum to a finer level of detail than has hitherto
been available.  As we have demonstrated, in this problem the details
are {\it crucial}: although the spectrum of the Harper equation at
\ecr~is a Cantor set, the gaps may be labeled  through a
topological invariant of orbits of the Harper map which is 
related to previously described rotation numbers for such systems
\cite{winding} and to the integrated density of states
\cite{numbertheory}. The ordering of the gaps depends on
number--theoretic properties of particular irrational frequency
$\omega$ \cite{slater,reorder}, while the gap indices determine the
exponents for the scaling of gap widths as a function of potential
strength.  The phase diagram for the Harper system will consist of
fractal tongues for all irrational frequencies $\omega$, and in the
tongues, the dynamics of the Harper map is on SNAs. The ubiquity of
such attractors and their correspondence with localized states
further underscores their importance \cite{ks,prss}.
\vskip0.5cm
{\sc Acknowledgment:} This research is supported by the Department of
Science and Technology, India. We have especially benefited from
correspondence with Jean Bellisard, and from discussions with Deepak
Kumar and Subir Sarkar.

\newpage
\section*{Figure Captions}

\vskip0.5cm

\noindent {\bf Figure 1:} The Integrated density of states (IDS)  (scale on the right)
and Lyapunov exponent ($\lambda$) (scale on the left) 
versus energy at \ecr. The gap labels $k$ are indicated for
the largest visible gaps. At every bifurcation, when $\lambda = 0$,
the dynamics is on a SNA. On the gaps, the IDS takes the constant
value $\Omega_k$ specified by Eq.~(\ref{winding}). \\
\ \\
{\bf Figure 2:}
(a) A strange nonchaotic attractor for \eps=2, $E=3$.
(b) The attractor for a value of $E$ corresponding to the gap $N=5$.
Note that the orbit has 5 branches that traverse the
range $-\infty < x < \infty$. \\
\ \\
{\bf Figure 3:} Ordering of the gaps for $\omega$ the golden mean. 
Only part of the Cayley tree described in the text is shown for 
clarity. Each node has two daughters except for 0, which has only one. \\
\ \\
{\bf Figure 4:} (a) Scaling of the gap widths, $w_N$ ($\bullet$), 
and depths $d_N$ ($\diamond$) as a function of gap index, $N$, at \eps~= \ecr. 
For clarity, the depths have been multiplied by a factor of 10. 
The dashed line fitting the depths has slope -1. The dotted lines
show the scaling of the two families of gaps; see the text for details.
(b) Scaling of the gap widths, $w_N$ for the largest few gaps as a
function of \eps~above \ecr. The solid lines are the power--laws
given in Eq.~(\ref{sc1}).\\
\ \\
{\bf Figure 5:} Phase diagram for the Harper map showing, below \ecr~(the dotted
vertical line) the regions of three--frequency quasiperiodic (Q)
orbits or extended states, 1--d attractors or gaps (G), and above
\ecr, regions of SNAs (S), and gaps (G). Only the largest gaps are
visible at this scale. All the gaps persist above \ecr, decreasing in
width according to Eq.~(\ref{sc1}), but the measure of the SNA region
(shaded) increases with \eps, as does the range of the spectrum.  

\begin{thebibliography}{99}
\bibitem{harper} P. G. Harper, Proc. Phys. Soc. London A {\bf 68},
74 (1955). 
\bibitem{hofstadter} D. R. Hofstadter, \prb{14}{2259}{1976}.
\bibitem{contexts} See \eg~ J. B. Sokoloff, Phys. Rep. {\bf 126}, 189
(1985); M. Y. Azbel, P. Bak, and P. M. Chaikin, \prl{59}{926}{1987};
I. I. Satija, \prb{49}{3391}{1994}; H.--J. St\"ockmann, {\it Quantum Chaos, an
introduction}, (Cambridge University Press, Cambridge, 1999), Chapter
4; P. B. Wiegmann, Prog. Theor. Phys. Suppl., {\bf 134}, 171 (1999);
Y. Last, in {\it XI Intern.\ Congress of Mathematical 
Physics (Paris, 1994)}, Ed.~D.~Iagolnitzer, 
Intern.\ Press, Cambridge, MA (1995), 
pp.~366--372; S.~Jitomirskaya, {\it ibid.}, pp. 373-382.
\bibitem{geisel} T. Geisel, R. Ketzmerick, and G. Petschel, 
\prl{66}{1651}{1991}; also in {\it Quantum Chaos}, edited by G.
Casati and B. Chirikov, (Cambridge University Press, Cambridge,
1995), pp. 633--60.
\bibitem{aubry} G. Andr\'e and S. Aubry, Ann. Isr. Phys. Soc., {\bf
3}, 133 (1980).
\bibitem{bellisard} J. Bellisard, R. Lima, and D. Testard,
\cmp{88}{107}{1983}. 
\bibitem{kohmoto} M. Kohmoto, \prl{51}{1198}{1983}; M. Kohmoto,
L. P. Kadanoff, and C. Tang, \prl{50}{1870}{1983}
\bibitem{pandit} S. Ostlund, R. Pandit, D. Rand, H. J. Schellnhuber,
and E. D. Siggia, \prl{50}{1873}{1983}.
\bibitem{op} S. Ostlund and R. Pandit, \prb{29}{1394}{1984}.
\bibitem{kspla}  J. Ketoja and I. I. Satija, \pla{194}{64}{1994}.
\bibitem{sk} J. Ketoja and I. I. Satija, \prl{75}{2762}{1995}.
\bibitem{kohmotoa} H. Hiramoto and M. Kohmoto,
Int. J. Mod. Phys. {\bf B6}, 281 (1992).
\bibitem{luck} J. -M. Luck, \prb{39}{5834}{1989}.
\bibitem{ks}  J. Ketoja and I. I. Satija, \physd{109}{70}{1997}.
\bibitem{prss} A. Prasad, R. Ramaswamy, I. I. Satija, and N. Shah,
\prl{83}{4530}{1999}.
\bibitem{gopy} C. Grebogi, E. Ott, S. Pelikan, and J. Yorke,
\physd{13}{261}{1984}; A. Prasad, S. S. Negi, and R.  Ramaswamy,
\ijbc{11}{291}{2001}.
\bibitem{bondeson} A. Bondeson, E. Ott, and T. M. Antonsen,
\prl{55}{2103}{1985}.
\bibitem{hof} For instance, the Hofstadter butterfly
\cite{hofstadter} can be trivially constructed as the 
set $\{(E,\omega): \lambda(E,\omega,\epsilon = 1) = 0\}$.
\bibitem{winding} The number $N$ of {\it complete}
traversals from $\tanh x = 1$ to $\tanh x = -1$ is related to the
winding number defined in R. Johnson and J.  Moser,
\cmp{84}{403}{1982} as $\Omega_N = M + N\omega$, $M, N$ integer.
See also F.  Delyon and B. Souillard, \cmp{89}{415}{1983}. 
\bibitem{numbertheory} J. Bellisard, in {\it From Number Theory to
Physics}, edited by M. Waldschmidt, P. Moussa, J.-M. Luck, and C.
Itzykson, (Springer Verlag, Berlin, 1992), pp. 538--630.  J.
Bellisard and B.  Simon, J. Funct. Anal., {\bf 48}, 408 (1982).
\bibitem{slater} N. B. Slater, Proc. Cambridge Philos. Soc. {\bf 63},
1115 (1967).
\bibitem{pbg} A. Pandey, O. Bohigas, and M.-J. Giannoni,
\jpa{22}{4083}{1989}. 
\bibitem{pr} A. Pandey and R. Ramaswamy, \pra{43}{4237}{1991}.
\bibitem{reorder} The ordering function, Eq.~(6) in \cite{pbg} or
the procedure leading to Eq.~(33) in \cite{slater} applies to 
numbers $y_j = \{j\omega\}$ and needs to be slightly modified for 
application to the present problem where all the $\Omega_k'$s are 
constrained to lie in the interval [1/2,1] since we are considering only
positive eigenvalues. The details are unwieldy but not very complicated, and
are given in S. S. Negi, Ph D. thesis, Jawaharlal Nehru University, 2001.
\bibitem{nr2} Subsidiary number--theoretic properties
help in arranging the remaining integers. If $m$ and $m^{\prime}$ are
daughters of $i_m$, then $(m+i_m)$ and $(m^{\prime} +i_m)$ are
consecutive Fibonacci numbers. Details of the overall scheme will
be given in a forthcoming publication.
\bibitem{wiegman} A. G. Abanov, J. C. Talstra, and P. B. Wiegmann,
Nucl. Phys. B {\bf 525}, 571 (1998).
\bibitem{experiments} C. Albrecht, J. H. Smet, K. von Klitzing, D.
Weiss, V. Umansky, and H. Schweizer, \prl{86}{147}{2001}.
\end{thebibliography}
\end{document}